\documentclass[prl,aps,twocolumn,showpacs,showkeys,superscriptaddress,groupedaddress]{revtex4}  
\usepackage{graphicx}  
\usepackage{dcolumn}   
\usepackage{bm}        
\usepackage{amssymb}   
\usepackage{color}

\begin{document}

\title{No hidden physics in resonance pole residue phase}

\author{S.~Ceci}
\email{sasa.ceci@irb.hr} \affiliation{Rudjer Bo\v{s}kovi\'{c} Institute, Bijeni\v{c}ka  54, HR-10000 Zagreb, Croatia}
\author{R.~Omerović} \affiliation{University of Tuzla, Urfeta Vejzagića 4, 75000 Tuzla, Bosnia and Herzegovina}
\author{H.~Osmanović} \affiliation{University of Tuzla, Urfeta Vejzagića 4, 75000 Tuzla, Bosnia and Herzegovina}
\author{M.~Uroić}
\affiliation{Rudjer Bo\v{s}kovi\'{c} Institute, Bijeni\v{c}ka  54, HR-10000 Zagreb, Croatia}
\author{B.~Zauner} \affiliation{Institute for Medical Research and Occupational Health, Ksaverska 2, HR-10000 Zagreb, Croatia}

\begin{abstract}
In hadron resonant scattering, there are four fundamental resonant parameters: real and imaginary part of the pole position, and the magnitude and the phase of the residue. Out of the four, the last one is the least understood. The search for the residue phase's physical meaning has focused on model-independent phases of the majority of the lowest-mass resonances. Here, we apply a simple mathematical identity to the amplitude in the complex plane to reveal the exact reason for the noticed regularity and show that there is no room for hidden physical variables in the residue phase. 
\end{abstract}


\maketitle


The defining property of a resonance phenomenon in a particle scattering is the first-order pole of the amplitude in the complex energy plane \cite{PDG}. Real part of the pole position is considered to be the resonance mass $M$, while its imaginary part is directly related to the total decay width $\Gamma$. The simplest relation for the resonant scattering amplitude is given by
\begin{equation}
    T=\frac{|r|\,e^{i\theta}}{M-E-i\,\Gamma/2}+T_B,\label{Eq:Amplitude}
\end{equation}
where $E$ is energy, $|r|\,e^{i\theta}$ is the pole residue, and $T_B$ is background amplitude. The magnitude of the residue $|r|$ is related to the strength of interaction between scattered particles, but also to the elasticity $2|r|/\Gamma$ in multichannel scattering. The physical meaning of the residue phase $\theta$ is still not clear, and it is the main topic of this Letter. 

Close to the pole, Eq.~(\ref{Eq:Amplitude}) with constant $T_B$ provides a good description for the resonant amplitude. Assuming elastic resonance, H\"ohler imposed unitarity on the amplitude and found that the residue phase $\theta$ is twice the phase of the background amplitude $T_B$ \cite{HohlerBible}. This implies that residue phase is probably not a fundamental property of the resonance since it does not come from the pole term, but from the background. Yet, background amplitude $T_B$ and resonant term are interconnected through unitarity; they are not independent. Moreover, the value of $\theta$ was completely free. This left room for physics.

The elastic residue phase in excited nucleon studies through pion-nucleon scattering was introduced and first time extracted by Cutkosky  \cite{Cutkosky}. Following in Cutkosky's footsteps, similar models were developed by Švarc \cite{Batinic1995, Batinic2010} (Zagreb model) and Vrana \cite{Vrana2000}. Alternative advanced analyses were performed and all pole parameters are extracted within the George Washington University model \cite{Arndt2006}, the Bonn-Gatchina model \cite{Anisovich2012}, the CBELSA/TABS collaboration analysis \cite{Sokhoyan2015}, and by the J\"ulich model \cite{Ronchen2015, Ronchen2022}. As statistics improved, the stability of the extracted parameters across models encouraged researchers to study their deeper meaning.

In a study of the $\pi N$ elastic cross sections at the elastic $\Delta(1232)$ resonances \cite{Ceci13}, using an amplitude similar to Eq.~(\ref{Eq:Amplitude}), it was noted that there is, in fact, a preferred value of $\theta$. It was confirmed for several nearly elastic resonances. This preferred value had something to do with the threshold: $\theta$ equaled twice the constant background phase that would reduce the amplitude to zero at the threshold. 

Few years later, a semi-empirical model was proposed \cite{Ceci17} with two distinct parameters. The first one was the aforementioned threshold phase. The other was the phase introduced earlier by Manley \cite{Manley} in an attempt to determine the pole position from the known Breit-Wigner mass $M_\mathrm{BW}$ and width. The residue phase was shown to be a sum of the two phases for numerous resonances.    

The same semi-empirical formula was simplified for elastic hadron resonances \cite{Ceci24}, such as the $\Delta(1232)$ and $f_0(500)$. In those cases, due to the unitarity, the threshold and Manley phases must be equal. This was confirmed by fitting the phase shift data close to the resonances: the constant background phase shift, related to the Manley's phase, would set the total phase shift to zero at the threshold. 

A recent study of the lightest resonances in partial waves for pion-pion, pion-nucleon, and kaon-nucleon scatterings have shown that this semi-empirical formula works rather well for around 2/3 of the first (lowest mass) resonances in partial waves \cite{Ceci25}. 

In this Letter we show that the success of the simple two-phase formula for the residue phase of the lightest resonances does not unveil a deeper physical meaning, but is merely a special case of a more general mathematical identity. The new formula works for most resonances, first and subsequent, and enables us to draw our main conclusion: even if there are some hidden physical variables in observed resonant parameters, there is simply no room for them in the residue phase.


A formula for the elastic residue phase of the lightest resonances in partial waves is introduced in Ref.~\cite{Ceci17} and confirmed in Refs.~\cite{Ceci24,Ceci25}  
\begin{equation}
    \theta = \alpha + \beta.\label{Eq:theta}
\end{equation}
Here 
\begin{equation}
    \alpha = - \arctan \frac{\Gamma/2}{M-E_0},\label{Eq:alpha}
\end{equation}
and
\begin{equation}
    \beta = - \arctan \frac{M_\mathrm{BW}-M}{\Gamma/2},\label{Eq:beta}
\end{equation}
with $E_0$ being a threshold energy, and $M_\mathrm{BW}$ the notorious (i.e.~model dependent) Breit-Wigner mass of the resonance. Threshold phase $\alpha$ is, thus, the angle at which the pole is seen from the threshold. Manley's phase $\beta$ is the angle at which the distance between pole mass $M$ and Breit-Wigner mass $M_\mathrm{BW}$ on the real axis is seen from the pole position.  

\begin{figure}[h!]
    \centering
    \includegraphics[width=0.47\textwidth]{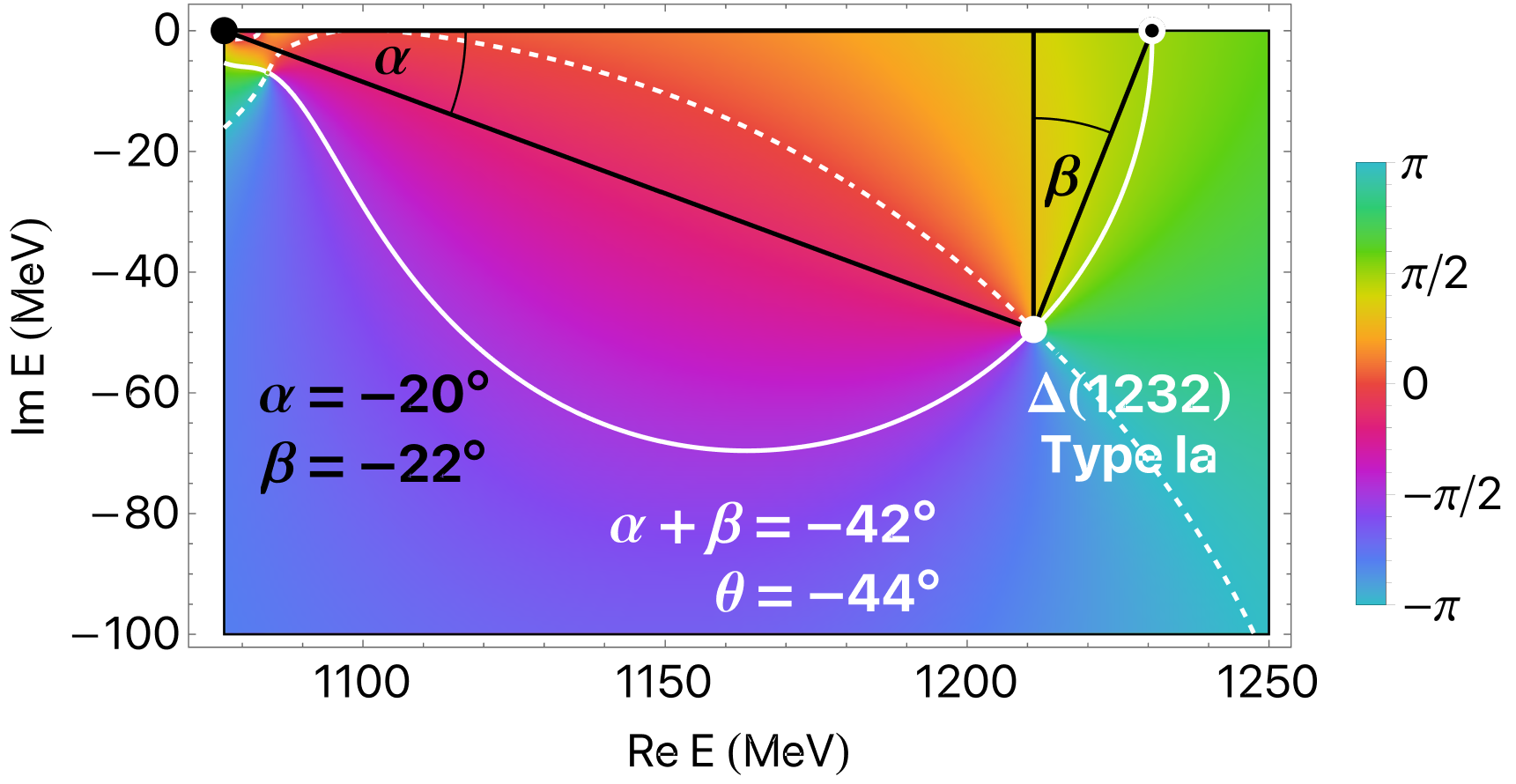}   
    \caption{Complex phase of the KH80 $\pi N$ scattering amplitude $T$ $(I=3/2,J^\pi=3/2^+)$ as a function of complex energy $E$ near the pole of $\Delta(1232)$. The pole is shown as the white circle, the threshold energy as the black circle, and the Breit-Wigner mass, which is here defined as the zero of the real part of the amplitude on the real axis, as the black and white circles. The true residue phase $\theta$ is determined numerically, while phases $\alpha$ and $\beta$ are calculated using Eqs.~(\ref{Eq:alpha}) and (\ref{Eq:beta}). The solid white line shows where $\mathrm{Re} \,T =0$, and the dashed one where  $\mathrm{Im} \,T =0$}
    \label{fig:firstkind}
\end{figure}

As an example how this relation works, we give the complex phase of the pion-nucleon scattering amplitude in the complex energy plane near $\Delta(1232)$ resonance. The amplitude we use is H\"ohler's KH80 \cite{HohlerPWA}, and to get to the non-physical Riemann sheet where resonance poles are situated, we use L+P analytic continuation devised by Švarc et al. \cite{LplusP}. We classify resonances satisfying Eq.~(\ref{Eq:theta}) as type Ia, and the result for one such prominent resonance is given in Fig.~\ref{fig:firstkind}.  

Eqs.~(\ref{Eq:alpha}) and (\ref{Eq:beta}), and consequently Eq.~(\ref{Eq:theta}) are inapplicable for resonances after the lightest, first one in each partial wave. We classify all these subsequent resonances as type II. In several waves, the formula fails even for the lightest resonances, which we name type Ib. Curiously, it was noted in Ref.~\cite{Ceci17} that some of these type Ib and type II resonances show a geometry similar to the one in these formulas. Angles analogous to $\alpha$ and $\beta$ could be easily defined deep in the complex plane, and the residue phase was the sum of the two new phases.

To explain this, we return to Eq.~(\ref{Eq:Amplitude}). Assuming we are close enough to the resonant pole, $T_B$ may be considered to be constant. Combining both terms into one, we get an amplitude with a simple zero at $Z$, simple pole at $P$, and overall constant complex factor $B=|B|\,e^{i\beta'}$ 
\begin{equation}
    T=\frac{E-Z}{P-E}\,B.\label{eq:general}
\end{equation}
Now we move into the complex plane and get to the zero $Z=E'_0-i\,\omega $. The real part of the zero, $E'_0$ is what the threshold $E_0$ was in our previous formulas, and the imaginary part $-\omega$ tells us how deep we need to go into the complex plane. Next we move to the right from the zero, parallel to the real axis, to find what was before the Breit-Wigner mass: the zero of the amplitude's real part $M'_\mathrm{BW}-i\,\omega$. Third interesting point for us is the pole, whose position we accordingly rewrite as $M-i\,\Gamma'/2-i\,\omega$. In this notation, the overall phase is given by a quite familiar relation
\begin{equation}
    \beta' = - \arctan \frac{M'_\mathrm{BW}-M}{\Gamma'/2}.\label{Eq:betaprime}
\end{equation}

To get the full residue phase, we need to calculate the contribution of the $E-Z$ at the pole position, let us call it $\alpha'$, which is given by another rather familiar relation
\begin{equation}
    \alpha' = - \arctan \frac{\Gamma'/2}{M-E'_0}.\label{Eq:alphaprime}
\end{equation}
The residue phase is now given by the sum of the two contributions 
\begin{equation}
    \theta = \alpha' + \beta'.\label{Eq:thetaprime}
\end{equation}

Hence, the simple semi-empirical physical formula from Refs.~\cite{Ceci17,Ceci24,Ceci25}, repeated here in Eq.~(\ref{Eq:theta}), is just a special case (when $\omega=0$) of a more general formula Eq.~(\ref{Eq:thetaprime}). Moreover, assuming $\omega=0$ and elastic resonance, and by imposing unitarity, one gets $\theta=2\alpha'=2\beta'$, which is the key formula of Ref.~\cite{Ceci24}.

In Eq.~(\ref{eq:general}) we assumed that the real part of the pole will be larger than the real part of the zero. For Ia type, this simply means that the resonance is above the threshold. Generally, this does not need to be true. In such cases, the residue phase must have an additional $\pm \,\pi$
\begin{equation}
\theta=\alpha'+\beta'\pm\,\pi.\label{Eq:thetaprimeprime}
\end{equation}


It has been shown in Ref.~\cite{Ceci25} that Eq.~(\ref{Eq:theta}) works for about 2/3 of the first nucleon and hyperon resonances, as well as for the non-strange mesons. Here, we begin by independently confirming its validity for prominent first resonances seen in partial waves of $\pi N$ elastic scattering. We use L+P study of KH80 \cite{LplusP} and compare it to L+P study of the single energy Wi08 SAID data from GWU \cite{Svarc13}. The results are shown in the first part of Table \ref{tab:one}. 

\begin{table}[h]
\begin{tabular}{cc|ccc|cc|c}
    \hline\hline 
     & & 
          \multicolumn{3}{c|}{KH80} &
          \multicolumn{2}{c|}{Wi08} &
          PDG\\
     & & $\frac{2|r|}{\Gamma}$ & $\theta$& $\theta_{OM}$   & $\theta$ & $\theta_{OM}$ & $\theta$\\
      Name & $J^\pi$ & (\%) & ($^\circ$) & ($^\circ$) & ($^\circ$) &  ($^\circ$) &  ($^\circ$)\\
    \hline
    \multicolumn{8}{c}{Type Ia: first resonances, Eq.~(\ref{Eq:theta})}\\
    \hline 
        $\Delta$(1232) & $3/2^+$ & 101  & $-44$ & ${\bf -42}$ & $-49$ & ${\bf -43}$ & $-46\pm2$\\ 
        N(1520) & $3/2^-$ & 57 & $-15$ & ${\bf -17}$ & $-7$ & ${\bf -10}$ & $-10\pm5$\\ 
        N(1675) & $5/2^-$ & 35 & $-23$ & ${\bf -20}$ & $-22$ & ${\bf -18}$ & $-22\pm 5$\\ 
        N(1680) & $5/2^+$ & 68 & $-16$ & ${\bf -16}$ & $-6$ & ${\bf -11}$ & $-20\pm10$\\  $\Delta$(1950) & $7/2^+$ & 39  & $-39$ & ${\bf -32}$ & $-24$ & ${\bf -24}$ & $-32\pm8$\\ 
        N(2190) & $7/2^-$ & 22 & $-17$ & ${\bf -25}$ & $-15$ & ${\bf -31 }$ & $0\pm30$\\        
    \hline
    \multicolumn{8}{c}{Type Ib: first resonances, Eqs.~(\ref{Eq:thetaprime}) or (\ref{Eq:thetaprimeprime}) }\\
    \hline 
        N(1440) & $1/2^+$ & 54 & $-87$ & ${\bf -85}$ & $-83$ & ${\bf -86}$ & $-90\pm10$\\ 
        $\Delta$(1620) & $1/2^-$ & 28 & $-104$ & ${\bf -104}$ & $-101$ & ${\bf -101}$ & $-100\pm20$\\    
        N(1720) & $3/2^+$ & 13  &  $-113$ & ${\bf -111}$ & $-86$ & ${\bf -86}$ & $-110\pm50$\\    
        $\Delta$(1930) & $5/2^-$ & 6 & $-38$ & ${\bf -38}$ & $-148$& ${\bf -152^*}$ & $-50\pm^{40}_{50}$\\    
        \hline
        \multicolumn{8}{c}{Type II: subsequent resonances, Eqs.~(\ref{Eq:thetaprime}) or (\ref{Eq:thetaprimeprime})} \\
        \hline
        $\Delta$(1600) & $3/2^+$ & 23 & $165$ & ${\bf 158^*}$ & $-128$ & ${\bf -135^*}$ & $-150\pm^{40}_{30}$\\   
        N(1710) & $1/2^+$ & 10  & $-108$ & ${\bf -108}$ & $139$ & ${\bf 139}$ & $-170\pm^{80}_{70}$\\ 
        N(1860) & $5/2^+$ & 6 & $-30$ & ${\bf -31}$ & $-74$ & ${\bf -85}$ & N/E\\    
        N(1895) & $1/2^-$ & 6 & $-110$ & ${\bf -111}$ & $102$ & ${\bf 102^*}$ & N/E\\ 
        N(1900) & $3/2^+$ & 5 & $-25$ & ${\bf -25}$ & N/O &-/- & $-10\pm30$\\ 
        $\Delta$(1900) & $1/2^-$ & 14 & $20$ & ${\bf 21}$ & $-51$ & ${\bf -51}$  & N/E\\ 
        $\Delta$(1910) & $1/2^+$ & 19 & $-83$ & ${\bf -82}$ & $179$ & ${\bf 179^*}$ & $-90\pm^{180}_{90}$\\   
        $\Delta$(1940) & $3/2^-$ & 9 & $148$ & ${\bf 148^*}$ & N/O &-/- & $-160\pm50$\\         
        N(2060) & $5/2^-$ & 10 & $-99$ & ${\bf -101}$ & N/O & -/- & $-110\pm20$\\   
    \hline \hline 
    \end{tabular}
    \caption{The $\pi N$ elastic residue phase $\theta$ determined for various excited nucleon states using L+P method \cite{LplusP} on H\"ohler's KH80 partial wave analysis \cite{HohlerPWA}, and GWU Wi08 single energy analysis \cite{Svarc13}. OM means Our Model, and values designated with $*$ are obtained using Eq.~(\ref{Eq:thetaprimeprime}). To get the idea how elastic are these resonances, we show $2|r|/\Gamma$ for KH80 as well. N/O means resonance Not Observed in analysis, and N/E that PDG has No Estimate for the residue phase. We find three basic types of excited nucleon states. Type Ia, the lightest resonances in each of partial wave that confirms validity of Eq.~(\ref{Eq:theta}). Type Ib are lightest resonances for which Eq.~(\ref{Eq:theta}) fails, but Eqs.~(\ref{Eq:thetaprime}) or (\ref{Eq:thetaprimeprime})
    work almost perfectly. Type II are subsequent resonances in partial waves. For them, Eqs.~(\ref{Eq:thetaprime}) or (\ref{Eq:thetaprimeprime}) work rather well}
    \label{tab:one}
\end{table}

\begin{figure}[h!]
    \centering
    \includegraphics[width=0.47\textwidth]{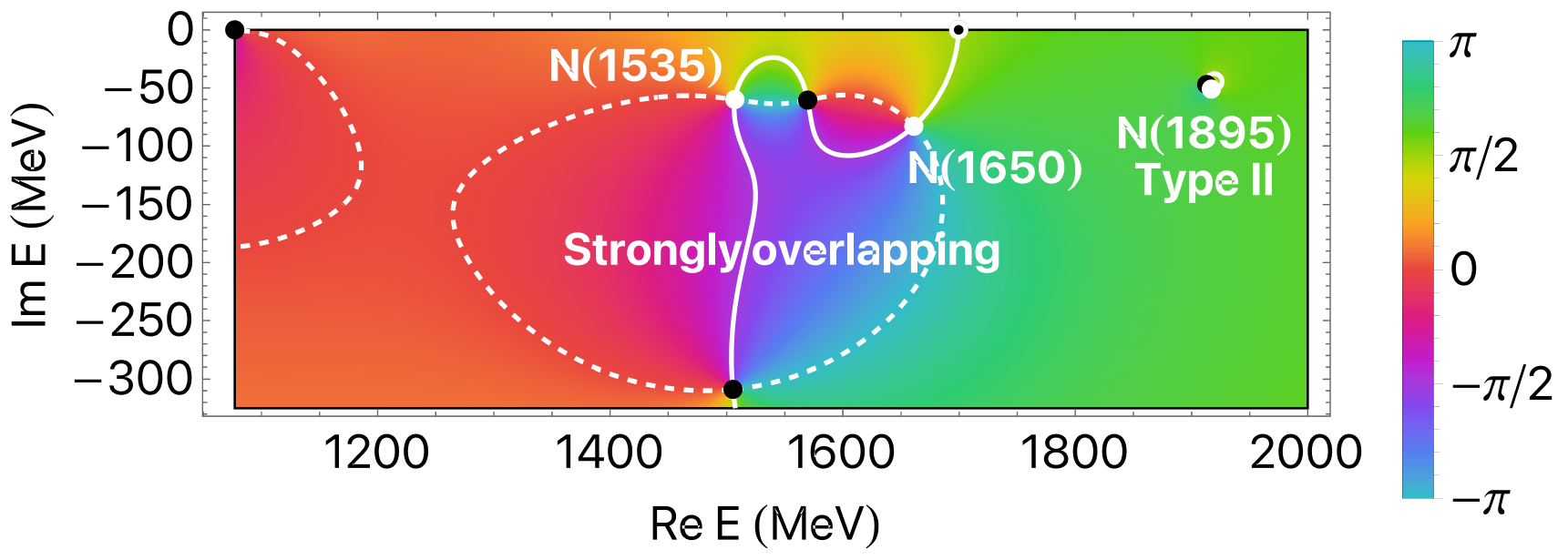}      
    \caption{The $\pi N$ elastic complex phase of the KH80 scattering amplitude $T$ ($I = 1/2$, $J^\pi=1/2^-$) as a function of complex energy $E$. Black circles are zeros, and white and black is the zero of the real part. We see strongly overlapping $N(1535)$ and $N(1650)$ resonances (white circles). As a result, $\mathrm{Re}\,T=0$ line (solid white line) does not cross the real axis, and without it, Eq.~(\ref{Eq:theta}) cannot be used for $N(1535)$. A multi-resonant model is needed here }
    \label{fig:zerothkind}
\end{figure}

Before we move on to our main result for the type Ib and type II resonances, we need to address several examples where our first resonance analysis was not successful. If there is strong overlap with a nearby resonance, as is the case of $N(1535)$ and $N(1650)$, we need a specific model to handle it, as was done in Ref.~\cite{Ceci17}. Similarly, if the line $\mathrm{Re}\,T=0$ (solid white line) does not cross the real axis, which also was the case for $N(1535)$, we cannot use this simple equation. Nevertheless, for this particular problem, one could use PDG estimate for the Breit-Wigner mass of $N(1535)$, as was done in Ref.~\cite{Ceci25}. The overlap is shown in Fig.~\ref{fig:zerothkind}, where we can also see the completely detached type II resonance N(1895).

The case of the N(2190) resonance is also interesting. Even though the discrepancy of Eq.~(\ref{Eq:theta}) results from KH80 and especially Wi08 in Table \ref{tab:one} are quite large, they are still within the PDG error estimates because the uncertainty estimate itself is very large. We have taken the PDG values and done the simplistic statistical averaging to get the result $-15^\circ\pm14^\circ$. If this were the official result, it would be in perfect accordance with both KH80 and Wi08, and justify calling N(2190) type Ia. The fact that Eq.~(\ref{Eq:theta}) produces substantial error for Wi08, and relatively large for KH80 tells us about the natural limit to this simple description: a strongly inelastic resonance very far from the threshold must depart somewhat from this model.

\begin{figure}[h!]   
    \includegraphics[width=0.47\textwidth]{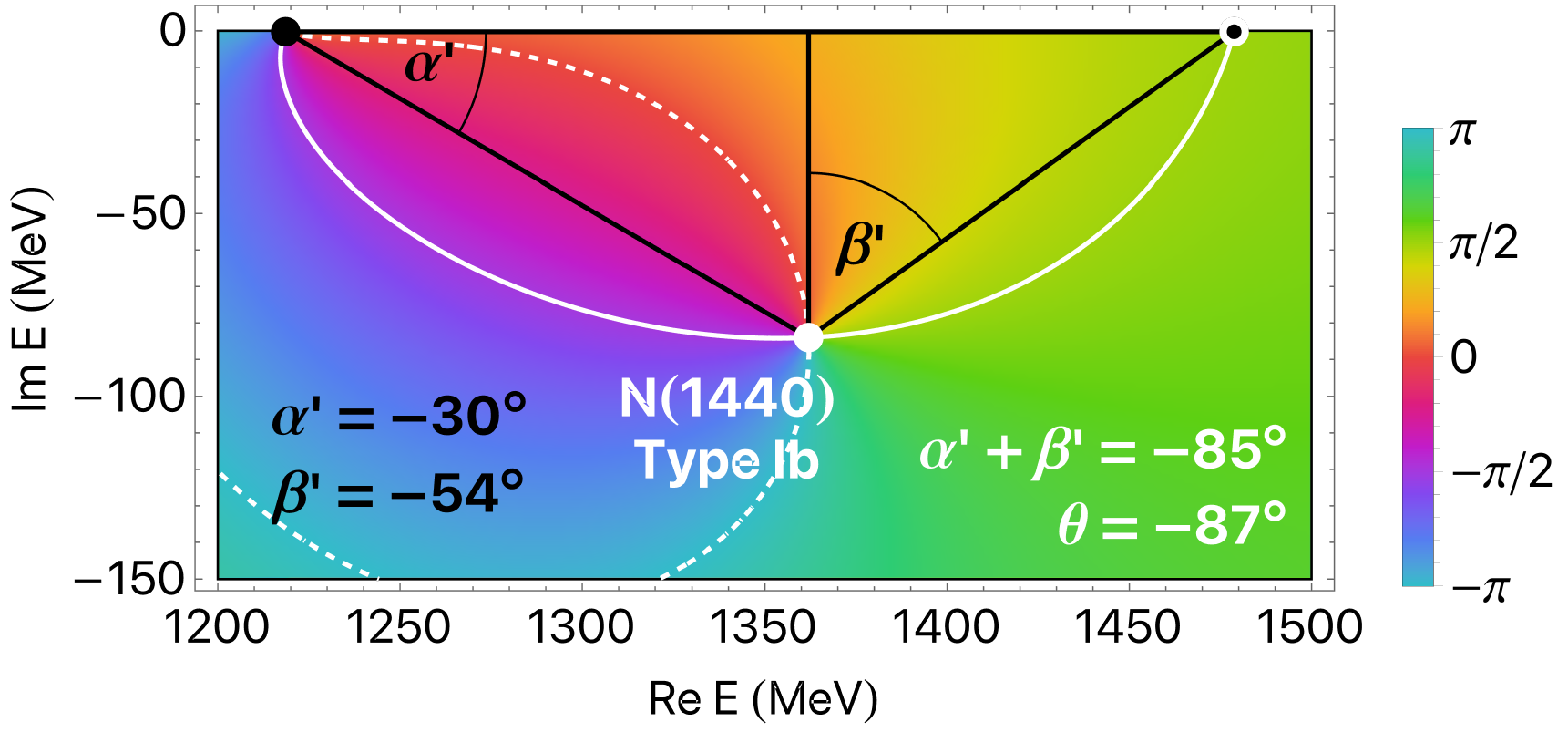}
    \includegraphics[width=0.47\textwidth]{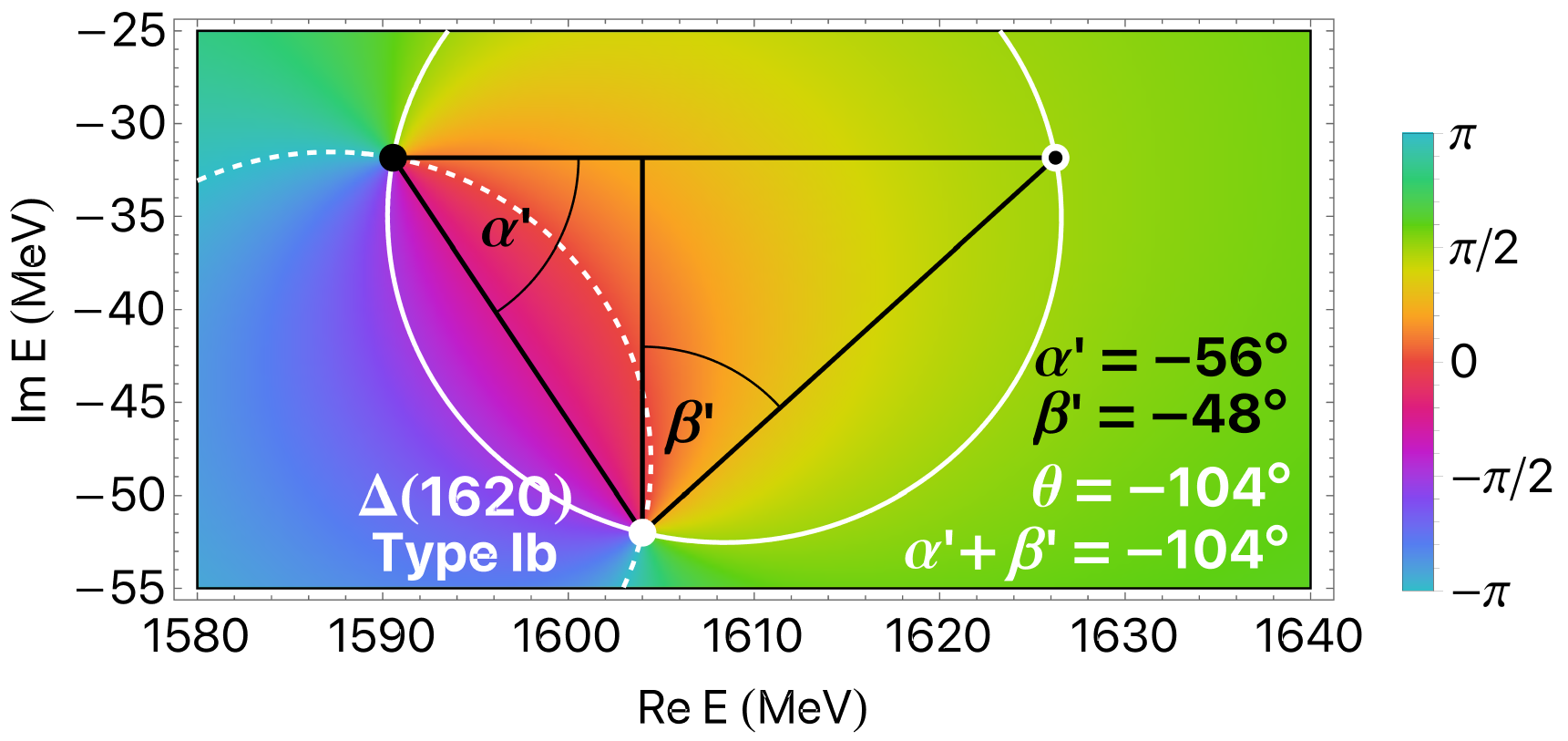}
    \includegraphics[width=0.47\textwidth]{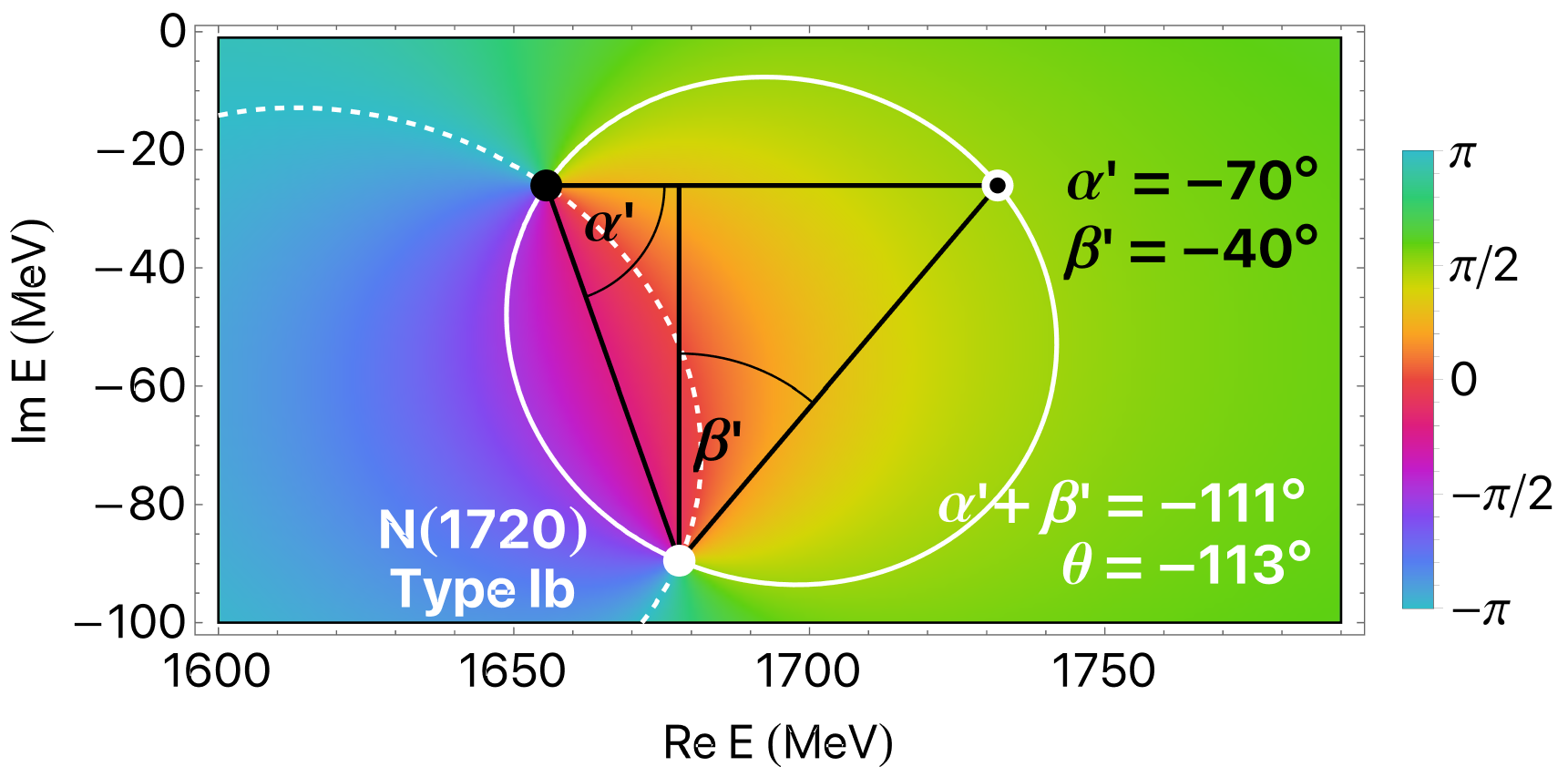}
    \caption{Complex phase of $T$ near resonances of the type Ib. Zero (black disk) is to the left, zero of the real part (black and white disk) is to the right, and the pole (white disk) is in between, similarly to the type Ia. N(1440) is almost type Ia, but it has a zero on the real axis, rather far from the elastic threshold. The other resonances are fully detached from the real axis, similar to type II. The two "zeros" are not too far from the real axis, and closer to the pole than in the case of Ia, resulting in $\alpha'$ and $\beta'$ being somewhat larger than Ia's $\alpha$ and $\beta$, which leads to $\theta$ typically around $-100^\circ\pm20^\circ$.}
    \label{fig:Ib}
\end{figure}

Moving on to the type Ib resonances. Four such notable cases are $N(1440)$, $\Delta(1620)$, $N(1720)$, and $\Delta(1930)$. All of them have a zero between the threshold and their poles. In the case of the Roper resonance, $N(1440)$, this zero is on the real axis. The main characteristic of all other type Ib resonances is that they are typically more inelastic than typical Ia resonance. A zero that is close to the real axis, and to the pole, produces larger $\alpha'$ than $\alpha$. In addition, $\beta'$ is also larger than $\beta$, which leads to a typical value of $\theta$ in the range $-100^\circ\pm20^\circ$. Interestingly, as the elasticity goes down, we note that the PDG error estimate gets larger. Again, the heaviest and most inelastic resonance of this type is somewhat different. The KH80 result is in accordance with PDG estimate, but the Wi08 result deviates significantly. This is more characteristic for type II resonances in Table \ref{tab:one}. Three typical Ib resonances are shown in Fig.~\ref{fig:Ib}.

In addition to type Ib resonances, we analyzed all other detached nucleon resonances for which we could find relevant parameters (pole, zero, and corresponding zero of the real part). For all of them numerical results are obtained using the original fit parameters from Refs.~\cite{LplusP,Svarc13}. We reconstructed their partial waves in the non-physical region, and numerically calculated needed parameters: zero, pole, zero of the real part having the same imaginary part as the zero, and the residue phase $\theta$ itself, which we then compare to the residue phase $\theta_\mathrm{OM}$ calculated from these parameters using Eqs.~(\ref{Eq:theta}), (\ref{Eq:thetaprime}), and (\ref{Eq:thetaprimeprime}). They all are also given in Table \ref{tab:one}. 

\begin{figure}[h!]
    \centering  
    \includegraphics[width=0.47\textwidth]{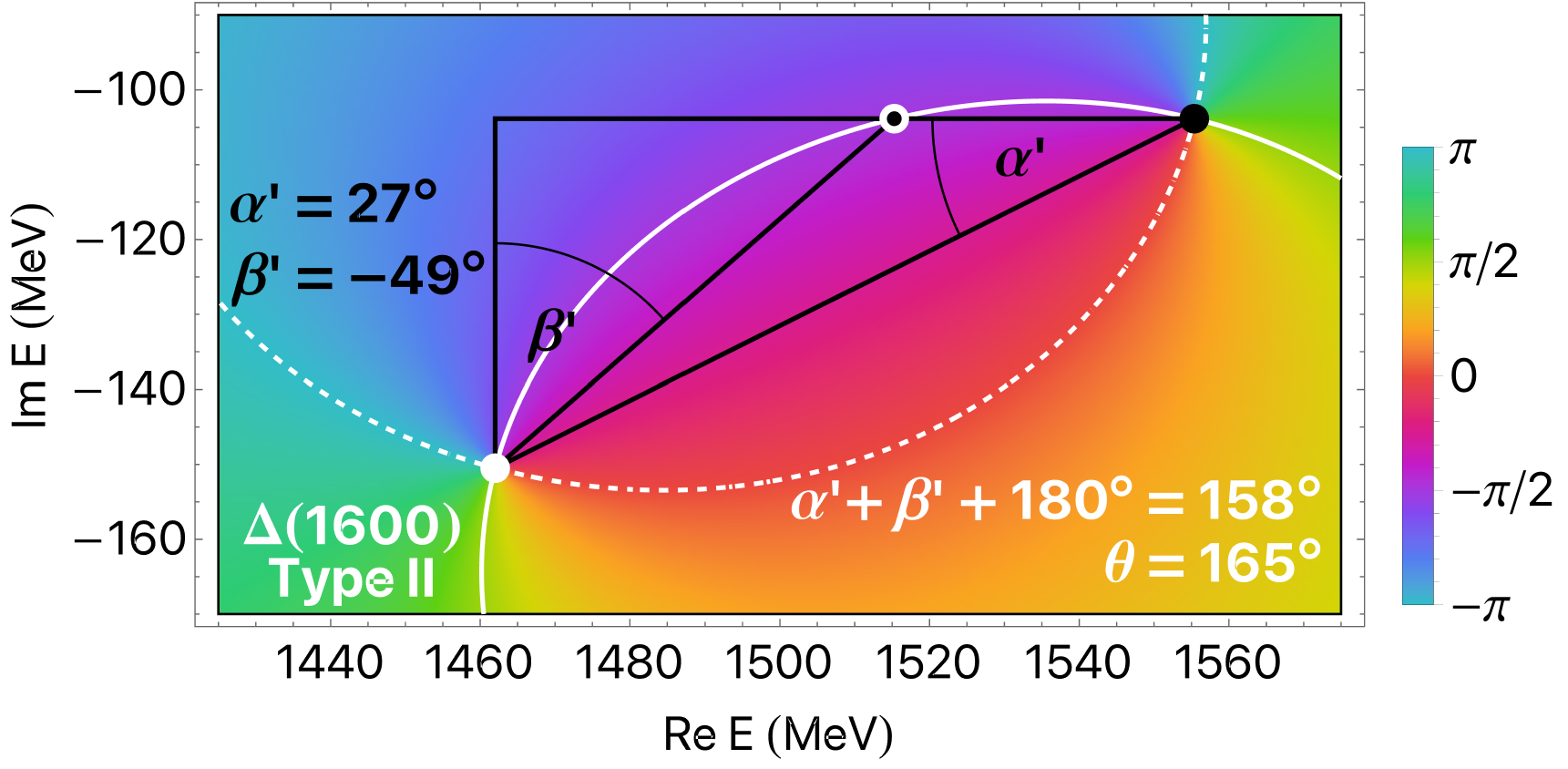}    \includegraphics[width=0.47\textwidth]{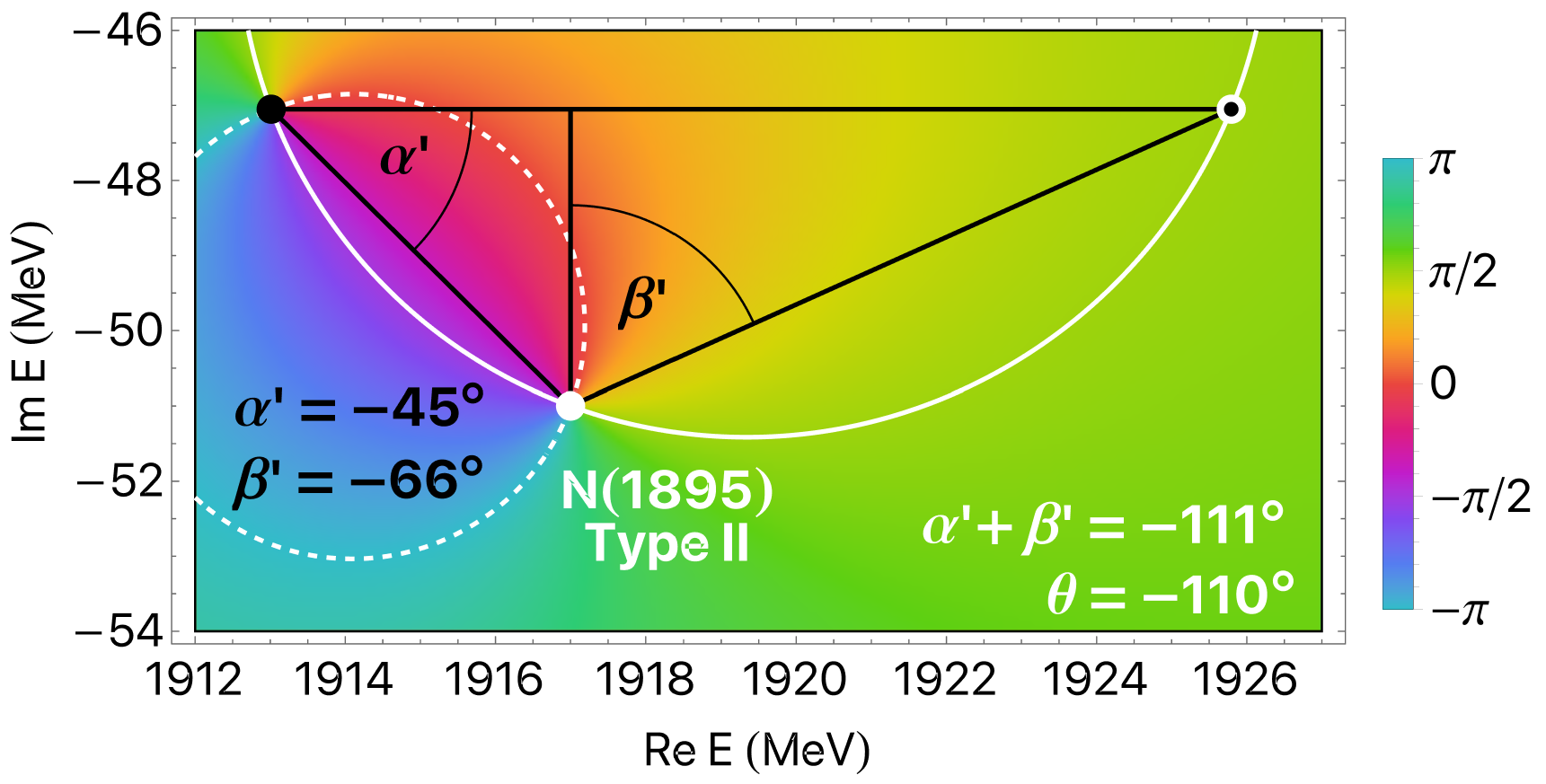}
    \includegraphics[width=0.47\textwidth]{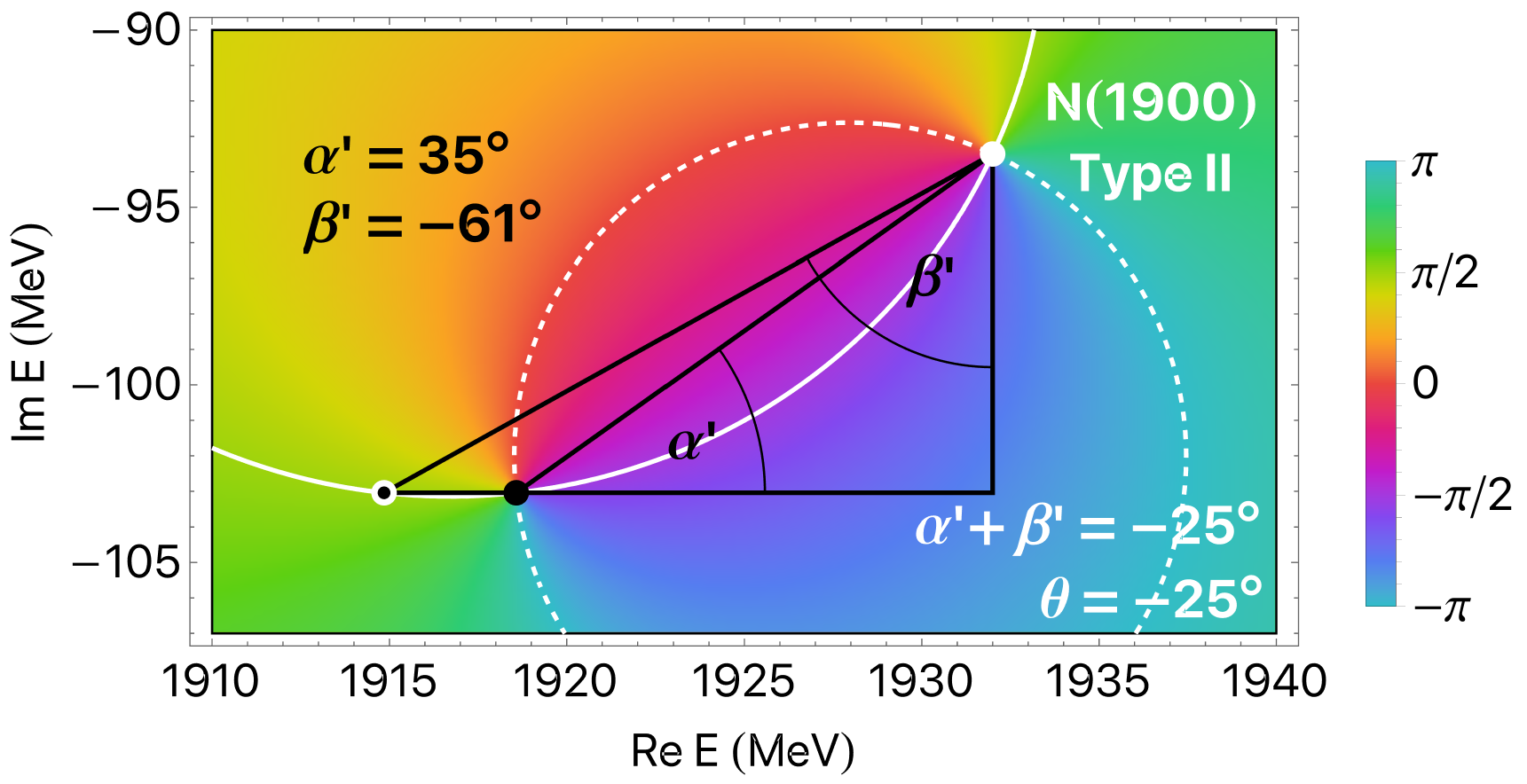}

    \caption{Complex phase of $T$ amplitude near the type II resonances. If the pole (white disk) is to the left of the zero (black disk), we must add or subtract $180^\circ$ according to Eq.~(\ref{Eq:thetaprimeprime}). N(1895) shows typical type Ib behavior, but it is not the first resonance; it is the third. Two other resonances show arrangements of the three characteristic points rarely seen in Ib type, and impossible for Ia. }
    \label{fig:II}
\end{figure}

It is clear from Table \ref{tab:one} that the difference between our model results (OM) for the KH80 and for Wi08 residue phases is almost insignificant from the original results for nearly all resonances. The discrepancy is typically much smaller than the PDG error estimate. The published L+P phases in Refs.~\cite{LplusP,Svarc13} are obtained by statistically averaging of several equivalent fits, similar to the ones that we use here. We show three illustrative examples of type II resonances in Fig.~\ref{fig:II}.

In conclusion, we have shown why the phase of type II resonances varies erratically among different analyses: model-dependent zero moves around the pole position, producing almost arbitrary residue phase. If there were any intrinsic resonant property hidden inside the phase, it would be impossible to extract it in any consistent way. What about the seemingly model-independent resonant types Ib and Ia? Most type Ib resonances have approximately the same residue phase (around $-100^\circ \pm 20^\circ$). It would be highly unusual to extract different physical information from the same number. And for type Ia, especially for nearly elastic resonances where $\beta\approx\alpha$, $\theta$ is determined purely from the threshold and the pole position. There is simply no room for any additional physical information to be hidden there.


\begin{thebibliography}{abcd1234}

\bibitem{PDG}
S.~Navas \textit{et al.} [Particle Data Group],
``Review of particle physics'',
Phys. Rev. D \textbf{110}, no.3, 030001 (2024)
doi:10.1103/PhysRevD.110.030001

\bibitem{HohlerBible} G.~H\"ohler, {\it Landolt-B\"ornstein:~Numerical Data and Functional Relationships in Science and Technology} Group~1, Volume~9, Subvolume~b, Part~2, pg.~202, (Springer-Verlag, 1983).

\bibitem{Cutkosky}  R.~E.~Cutkosky, C.~P.~Forsyth, R.~E.~Hendrick, and R.~L.~Kelly, \textit{et al.}: ``Proc. 4th Conf. on Baryon Resonances in Toronto'', ed. N. Isgur, p. 19 (1980).

\bibitem{Batinic1995} M.~Batini\' c, I.~\v Slaus, A.~\v Svarc, and B.~M.~K.~Nefkens, Phys.~Rev.~{\bf C 51}, 2310 (1995). 

\bibitem{Batinic2010} M.~Batini\' c, S.~Ceci, A.~\v Svarc, and B. Zauner, Phys.~Rev.~{\bf C 82}, 038203 (2010).

\bibitem{Vrana2000}
T.~P.~Vrana, S.~A.~Dytman and T.~S.~H.~Lee,
Phys. Rept. \textbf{328}, 181-236 (2000)
doi:10.1016/S0370-1573(99)00108-8
[arXiv:nucl-th/9910012 [nucl-th]].

\bibitem{Arndt2006} R.~A.~Arndt, W.~J.~Briscoe, I.~I.~Strakovsky, and R.~L.~Workman, Phys. Rev. {\bf C 74} 045205 (2006).

\bibitem{Anisovich2012}
A.~V.~Anisovich, R.~Beck, E.~Klempt, V.~A.~Nikonov, A.~V.~Sarantsev and U.~Thoma,
Eur. Phys. J. A \textbf{48}, 15 (2012)
doi:10.1140/epja/i2012-12015-8
[arXiv:1112.4937 [hep-ph]].

\bibitem{Sokhoyan2015}
V.~Sokhoyan \textit{et al.} [CBELSA/TAPS],
Eur. Phys. J. A \textbf{51}, no.8, 95 (2015)
[erratum: Eur. Phys. J. A \textbf{51}, no.12, 187 (2015)]
doi:10.1140/epja/i2015-15187-7
[arXiv:1507.02488 [nucl-ex]].

\bibitem{Ronchen2015}
D.~R\"onchen, M.~D\"oring, H.~Haberzettl, J.~Haidenbauer, U.~G.~Mei\ss{}ner and K.~Nakayama,
Eur. Phys. J. A \textbf{51}, no.6, 70 (2015)
doi:10.1140/epja/i2015-15070-7
[arXiv:1504.01643 [nucl-th]].

\bibitem{Ronchen2022}
D.~R\"onchen, M.~D\"oring, U.~G.~Mei\ss{}ner and C.~W.~Shen,
Eur. Phys. J. A \textbf{58}, no.11, 229 (2022)
doi:10.1140/epja/s10050-022-00852-1
[arXiv:2208.00089 [nucl-th]].

\bibitem{Ceci13} S. Ceci, M. Korolija, and B. Zauner
Phys. Rev. Lett. {\bf 111}, 112004 (2013).

\bibitem{Ceci17} S.~Ceci, M.~Had\v zimehmedovi\' c, H.~Osmanovi\' c, A.~Percan, and B.~Zauner, 	Sci.~Rep.~{\bf 7}, 45246 (2017). 

\bibitem{Manley} D.~M.~Manley, ``Masses and widths of $N$ and $\Delta$ resonances'', Phys. Rev. D \textbf{51}, 4837 (1995)

\bibitem{Ceci24} S. Ceci, M. Vukšić, and B. Zauner, arXiv:2005.11564 [hep-ph]

\bibitem{Ceci25} S. Ceci, H. Osmanović, and B. Zauner, 	arXiv:2505.16880 [hep-ph]

\bibitem{LplusP}
A.~{\v{S}}varc, M.~Had{\v{z}}imehmedovi{\'c}, R.~Omerovi{\'c}, H.~Osmanovi{\'c} and J.~Stahov,
Phys. Rev. C \textbf{89}, 045205 (2014)
doi:10.1103/PhysRevC.89.045205
[arXiv:1401.1947 [nucl-th]].

\bibitem{HohlerPWA}
G.~Hohler, F.~Kaiser, R.~Koch and E.~Pietarinen,
Phys. Daten \textbf{12N1}, 1 (1979)

\bibitem{Svarc13} 
A.~Švarc, M.~Hadžimehmedović, H.~Osmanović, J.~Stahov, L.~Tiator, and R.~L.~Workman, 
Phys. Rev. C \textbf{88}, 035206 (2013)
doi:10.1103/PhysRevC.88.035206


\end{thebibliography}
\end{document}